\documentstyle[prl,aps,graphicx]{revtex}

\newcommand{\be}{\begin{equation}} \newcommand{\ee}{\end{equation}}
\newcommand{\ba}{\begin{array}{l}} \newcommand{\ea}{\end{array}}
\newcommand{\al}{\alpha} 
\newcommand{\roroc}{{\rho \over \rho_c}}
\begin{document} \draft 
\title{Rotation, mass loss and pulsations of the star:\\
an analytical model}
\author {Zakir F.Seidov}  
\address{Department of Physics, Ben-Gurion University, Beer-Sheva 84105,
Israel\\ email:seidov@bgumail.bgu.ac.il} \maketitle 
\begin{abstract} The characteristics  of the  star model  with the 
"prescribed" density distribution 
$\rho=\rho_c [1-(r/R)^\alpha]$ are analytically studied. 
The model has been first considered briefly in our 30-year old note of a
restricted circulation \cite{pseu}. 
\end{abstract}
\subsection*{Introduction}
We choose the distribution of density $\rho (r)$ in a spherical-symmetric star in the form
\be \label{denlaw}\rho (r)=\rho_c (1-x^\al),\quad x={r\over R}; \ee
here $\rho_c=\rho(0)$ is the central density, $r$ is a running radius, $R$ is the radius 
of the model, $\al$ is a free parameter, ($\al\ge\,0$, as we require 
$d \,\rho/d\,r\le\,0$).
\subsection*{Central-to-mean density ratio}
From Eq. (\ref{denlaw}) we have for a cumulative mass $m$, total mass $M$ and for the
central-to-mean density ratio $\rho_c/\overline \rho$:
\be \ba  
m(x)= 4 \pi \rho_c R^3 \biggl({x^3\over 3}-{x^{3+\al}\over
{3+\al}}\biggr);\quad M=m(1)={4\, \pi\over 3}\rho_c\,R^3\,{\al\over
3+\al}= {4\,\pi\over 3}\overline\rho\,R^3;\quad 
{\rho_c\over \overline \rho}={3+\al\over \al}.\ea \ee
\subsection*{Central moment of inertia}
The moment of inertia of the star about its center is:
\be I=\int\nolimits_0^M\,r^2\,dm={4\,\pi\over 5} \rho_c\,R^5 {\al\over 
5+\al}:\ee 
with help of Eq. (2) we may write down $I$ in terms of $M$ and $R$:
\be I={3\over 5}{3+\al\over 5+\al}\,M\,R^2.\ee

\subsection*{Gravitational potential energy }

The total potential energy of the star is:
\be \ba W=-G\,\int\nolimits_0^M\,{m\over r}\,dm=
-{16\,\pi^2\over 15}{{\,{\al^2}\,\left( 11 + 2\,\al \right) \,
\over{( 3 + \al ) \, \left( 5 + \al \right) \,
     \left( 5 + 2\,\al \right)}}G\,{R^5}\,{\rho_c}^2}=
{{-3\,\left( 3 + \al \right) \,
     \left( 11 + 2\,\al \right)}\over 
   {5\,\left( 5 + \al \right) \,
     \left( 5 + 2\,\al \right) }}{ \,G\,M^2\over R} .\ea\ee

\subsection*{WUM-ratio}

The central gravitational potential $U(0)$ and the potential at the surface $U(R)$ 
of the spherically-symmetric star are 
\be U(0)=-4\,\pi\,G\,\int_0^R \rho(r)\,r\,dr;\quad U(R)=-{G\,M\over R}.\ee
for the model (\ref{denlaw}):
\be U(0)={{-2\,\al\,G\,\pi \,{R^2}\,{\rho_c}}\over {2 + \al}}=
{3\over 2}{3+\al\over 2+\al}U(R). \ee
Combining Eqs (2), (5), (7) we obtain the $WUM$-ratio \cite{WUM} for our model:
\be WUM={W\over U(0)\,M}={2\,(2 + \al)\,(11 + 2\,\al)\over 5\,(5 + \al)\,(5 +
2\,\al)}.\ee

\subsection*{Pressure-density relation}

Now we find the relation between pressure $P$ and density $\rho$ in the
star, which has the density distribution (\ref{denlaw}) at hydrostatic equilibrium.
In other words, we find the equation of state of matter which leads to
distribution of density (\ref{denlaw}) in the star. We remind that a star with the
equation of state $P(\rho)$ which does not include a temperature is called
pseudopolytrope \cite{Krat}.\\
The equation of hydrostatic equilibrium of spherically-symmetric star
reads:
\be \label{hesss} {1\over \rho}{d\,P\over d\,r}=-G\,{m\over r^2}.\ee 
Using Eqs (1, 2) and integrating Eq. (\ref{hesss}) with initial condition $\,P(0)=P_c\,$
(central
pressure) we obtain the distribution of pressure and density within the star:
\be\label{Prox}\ba P(x)=P_c - G\,\pi \,{R^2}\,{{{\rho_c}}^2}\, \left[ {{2\,{x^2}}\over 3}
-      {{4\,\left( 6 + \al \right) \,{x^{2 + \al}}}\over 
  {3\,\left( 2 + \al \right) \,\left( 3 + \al \right) }} + {{2\,{x^{2 + 2\,\al}}}\over 
       {\left( 3 + \al \right) \,\left( 1 + \al \right) }} \right];\\
\rho(x)=\rho_c (1-x^\al); \quad x=r/R.\ea \ee

 From Eq. (\ref{Prox}) using condition $P(1)=0,\,$ we obtain the
following relations for the central pressure:

\be \label{Pc} P_c=
{{2\,\pi\,{\al^2}\,\left( 4 + \al \right)}\over {3\,\left( 1 +
 \al \right) \,\left( 2 + \al \right) \,
     \left( 3 + \al \right) }}\,G \,{R^2}\,{{{\rho_c}}^2}=
{{3\,\left( 3 + \al \right) \,\left( 4 + \al \right)
}\over{8\,\pi\,\left( 1 + \al \right) \,\left( 2 +
 \al \right)  \,}}{ \,G\, M^2\over {R^4}}.\ee

From Eqs. (\ref{Prox}, \ref{Pc}) we obtain the following relation
between $P$ and $\rho$ ("equation of state", EOS):

\be \label{EOS} \ba {P\over P_c}= 1 + (1-\roroc)^{2/ \al}\,\left[ 1 + {2\over \al}\,\roroc
+ {3\,( 2 + \al ) \over \al^2\,( 4 + \al )}\,(\roroc)^2 \right]. \ea\ee

Note that for any given pair of parameters $\rho_c,\,P_c$, 
(and {\it fixed} $\al$) we have a particular EOS given
by Eq. (12). Therefore these particular pseudopolytropes have the two-parametric
 EOS, while e.g. the classic polytropes with given adiabatic index $\gamma$
have one-parametric EOS: $P=K\,\rho^\gamma$, the parameter $K$ being related with
polytropic temperature, see for more details \cite{Chan}.  
If we calculate a model with other values of central
density $\rho_c^*$ and pressure $P_c^*$ 
with the same EOS (12), the resulted distribution of
density will not coincide with the law (\ref{denlaw}, see the next Section).
Particular cases $\al=1$ and $\al=2$ give the more simple equations of state:

\be \label{EOS1} {P\over P_c}={1\over 5}\,(\roroc)^2 \,\biggl(6  + 8\,\roroc - 
     9\,(\roroc)^2\biggr);\quad \al=1;\ee
\be \label{EOS2} {P\over P_c}={1\over 2}(\roroc)^2\,(1 + \roroc); \quad \al=2.\ee

\begin{figure} \includegraphics[scale=.6]{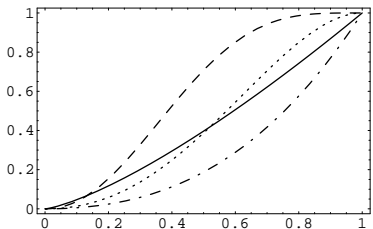}
\caption{Equations of state for different values of $\al$: 
 .... - $\al =1$, -.-.  - $\al =2$, - - -  - $\al =1/2$, and 
solid line -  polytropic EOS with $\gamma =4/3$. Abscissae are values of density
$\roroc$, ordinates are values of pressure $P/P_c$.}
\end{figure}

Although these equations of state, and the more general EOS (\ref{EOS}) are derived
and are formally valid only for $\roroc\le\,1$, we will loosely use them also at larger
densities,  see the next Section.

\subsection*{Density distribution in the $\al=2$ model}

Here we calculate density distribution of the particular model with $\al=2$,
which corresponds to EOS given by  Eq. (\ref{EOS2}). With this EOS,
the equation of hydrostatic equilibrium (\ref{hesss}) is reduced to the next equation for
the density distribution $\rho (x)$:

\be\label{hesss2}{d\over d\,x}[x^2\,({3\over 2}\,\rho\,+1){d\,\rho\over
d\,x}]=-15\,x^2\,\rho.\ee

In deriving this equation, we used the first relation in Eq. (11). Important notice
is that Eq. (\ref{hesss2}), at the case $\rho(0)=1$, has the solution which corresponds to
the law  (\ref{denlaw}) (with $\al=2$), and all parameters
corresponding to this case are taken to be equal to unit (that is, at $\rho(0)=1,\,$
we have $R=1,\,\,$ $M=1\,$ and so on. Also, the dimensionless running radius in Eq.
(\ref{hesss2}) $x$ is expressed in units of the total radius of star $R$, which is 
{\it not} equal to $1$ at $\rho(0)\neq 1$. 

\begin{figure} \label{alpha2} \includegraphics[scale=.6]{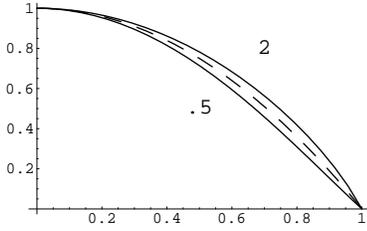}
\caption{Density distribution for EOS $P={\rho^2\,(\rho+1)\over 2}$ at three values
of central density $\rho(0)=2,\, 1$ and $1/2$. Abscissae are values of dimensionless
running radius $x=r/R$, ordinates are values of ${\rho(x)\over \rho(0)}$.}
\end{figure}

Eq. (\ref{hesss2}) was solved numerically for two
values of $\rho(0)$, $2$ and 
$1/2$. In these two cases the radius of the star is $1.169156$ and $.9070496$, 
respectively.
The resulting density distribution is shown in Fig. 2, where abscissae are $x=r/R$
and ordinates are $\rho(x)/\rho(0)$; $\rho(0)=2$ case corresponds to
the upper solid curve, while $\rho(0)=1/2$ corresponds to the lower
solid line. Also
shown is the $standard$ distribution - Eq. (\ref{denlaw})
with $\al=2$, dash line in Fig. 2. For values of the mean-to-central
density ratio 
$\overline\rho/\rho(0)$ we have: $.363325$, $.4$ and $.442107$ at
$\rho(0)$ equal to $1/2$, $1$ and $2$, respectively. For larger central
density,
EOS (\ref{EOS2}) has larger adiabatic index $\gamma$, EOS is more stiff,
and this leads to more homogeneous configuration with larger
mean-to-central density ratio. 

\subsection*{Pade approximants}

We briefly describe the Pade approximants method which we  use 
further in this paper.\\

From the segment of the series:

\be s_4=1+\sum_{k=1}^{k=4}\,a_k\,x^k,\ee

coefficients of Pade(2,2) approximants \cite{Baker}

\be PA= {1+A\,x +B\,x^2 \over 1+ C\,x+D\,x^2}\ee

are defined as follows \cite{McCrack}:

\be  D={a_3^2-a_2\,a_4\over 
a_2^2-a_1\,a^3};\,\,
C={a_1\,a_4-a_2\,a_3\over a_2^2-a_1\,a^3};\,\,
B=a_2+a_1\,C+D;\,\,A=a_1+C.\ee

\subsection*{Pade approximants to density distribution} 

From Eq. (\ref{hesss2}) we have series expansion at the center:

\be \ba \rho(x)=\rho(0)\,[1+\sum_{k=1}^{k=4}\,a_k\,x^{2\,k}];\quad
a_1=-{5\over 2 + 3\,\rho(0)};\quad 
a_2={{-15\,\left( -1 + \rho(0) \right) }\over 
   {{{\left( 2 + 3\,\rho(0)  \right) }^3}}};\\ \\

a_3={{150\,\left( 1 - 9\,\rho(0)  \right) \,\left( -1 + 
       \rho(0) \right) }\over{7\,{{\left( 2 + 
          3\,\rho(0) \right) }^5}}};\quad

a_4={{25\,\left( -1 + \rho(0) \right) \,\left(-5 + 222\,\rho(0)  - 
837\,{{\rho(0)}^2}\right)}\over{7\,{{( 2 + 3\,\rho(0))}^7}}}.
\ea\ee

Evidently, at $\rho(0)=1$ the series is reduced to $\rho(x)=1-x^2$.
Expressions for coefficients of Pade(2,2) approximants are too
cumbersome at the
general case of arbitrary $\rho(0)$ so we present only one particular case:

\be \label{rhopade} \rho(0)=2:\quad \rho_{Pade}(x)=2\,{{12816384 -
13840960\,{x^2} + 
     3314065\,{x^4}}\over 
   {12816384 - 5830720\,{x^2} + 
     45345\,{x^4}}}.\ee

Using this "analytic solution" we find the radius of the model equal to
$1.17720$ which is $1.0068$ times the calculated value $1.169156$.
Also, we calculated mean-to-central density ratio according to 
(\ref{rhopade}) and found value $0.435602$ which is
$0.98528681$ of the numerical value $.442107$.

  \subsection*{Mean adiabatic index} The adiabatic index is defined as:

\be\gamma={\,d\,\ln\,P\over d\,\ln\,\rho}.\ee
Distribution of adiabatic index within the model is:
\be \gamma (x)={{2\,( 1 + \al ) \,( 2 + \al ) \,{x^{2 - \al}}\,
     ( 3 + \al - 3\,{x^\al}) \,{{(1 - x^\al)
}^2}}\over {\al\,\left[ 4\,{\al^2} + {\al^3} - (6 + 11\,\al + 
       6\,{\al^2} + {\al^3})\,{x^2} + (12+ 
       14\,\al + 2\,{\al^2})\,{x^{2 + \al}} - (6+  3\,\al)\,{x^{2 +
2\,\al}} \right] }}.\ee
At the surface of the star (at $x=1$) $\gamma=2$ for any $\al$ while the value of
$\gamma$  at the center of the star depends on $\al$: 
$\gamma(0)=0$  at $\al<2$, $\gamma(0)=\infty$ at $\al>2$, and
$\gamma(0)=5/2$ at $\alpha=2$.\\  
Moreover the function $\gamma (x)$ is non-monotonic at $\al<2$, while
at $\al \ge 2$ the function $\gamma(x)$ is monotonically decreasing
with increasing $x$, relative running radius, see Fig. 3.

\begin{figure} \includegraphics[scale=.6]{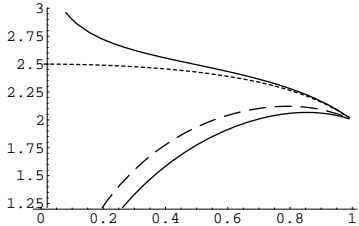} \label{gamma}
\caption{Adiabatic index as function of $x=r/R$, according to Eq. (22), 
curves from the lowest one to the 
upper one correspond to values of $\al=1/2,\,1,\,2,\,2.1$, respectively. }
\end{figure}

Therefore at $\al<2$, in the inner parts of the star, the value of  $\gamma$ is  
less than $4/3$, the critical value of adiabatic index, while in the outer regions  
$\gamma$ is always larger than $4/3$.\\ 
The stability of the
star against radial perturbations is defined by {\it the mean value of 
adiabatic index}.  
That is why it is of  interest to calculate the mean adiabatic index,
$\overline\gamma$, which is defined as follows: 

\be \overline\gamma=
{1\over \int\nolimits_V\,P\,dv}\int\nolimits_V\,\gamma\,P\,dv;\ee

here an integration is over the volume of star. We obtain at $\al<5$:

\be  \overline\gamma
={12\,(7+\al)\over (5-\al)(11+2\,\al)}.\ee

For any $\al\ge 5$,  $\overline\gamma\rightarrow\infty$ due to divergence of
$\gamma(r)\,P(r)\,r^2$ at the center of the star.  
The function $\overline\gamma (\al)$ is monotonically increasing and even
at $\al=0$, $\overline\gamma =84/55>4/3$, that is the model is always stable
against to the radial perturbations.

\subsection*{Ellipticity of slow rotating star}

For the slow rotation, the ellipticity distribution inside the spherically-symmetric star
is governed by the Clairaut's equation \cite{C1743}:

\be \label{Clair}
[{e''(x)\over x^2}-{6\,e(x)\over x^4}]\,\int_0^x\,\rho(t)\,t^2\,dt+2\,\rho(x)
[{e(x)\over x}+e'(x)]=0.\ee

Here $x=r/R$ is dimensionless running radius, $0\le x \le 1$, and $e(x)$ is
an ellipticity of the equidensity surfaces within the star, 
$e(x)=1-r_p(x)/r_{eq}$, with $r_p$ and $r_{eq}$ being the polar and equatorial
radii of the equidensity surfaces which are assumed to be the biaxial
ellipsoids of revolution.
 
The Equation (\ref{Clair}) can be solved in terms of the series expansion 
at the center, which we write down here for two cases:

\be \ba \al =1:\quad
e(x)=\sum_{i=0}^\infty c_i\,x^i;\quad c_i={3\over 4}{i^2+5\,i-4\over
i\,(i+5)}\,c_{i-1}; \quad i\ge 1;\\ \\ 
e(x)=1 + {x\over 4} + {{15\,{x^2}}\over {112}} + 
  {{75\,{x^3}}\over {896}} + {{25\,{x^4}}\over {448}} + 
  {{69\,{x^5}}\over {1792}};\\ \\
\al=2:\quad
e(x)=\sum_{i=0}	^\infty c_i\,x^{2\,i};\quad c_{i+1}={3\over 5}{2\,i^2+
9\,i+2\over
(i+1)\,(2\,i+7)}\,c_{i-1}\quad i\ge 0;\\ \\
e(x)=
1 + {{6\,{x^2}}\over {35}} + {{13\,{x^4}}\over {175}} + 
  {{52\,{x^6}}\over {1375}} + {{141\,{x^8}}\over {6875}} + 
  {{1974\,{x^{10}}}\over {171875}}. \ea\ee

We solved Eq. (\ref{Clair}) numerically for the model (1) with two values
of $\al$, see Figs. 4 and 5. 

\begin{figure} \includegraphics[scale=.6]{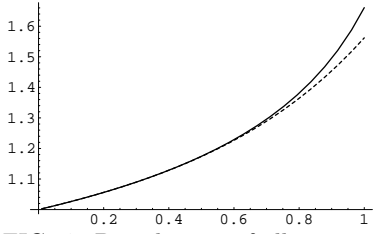}
\caption{Distribution of ellipticity within the star for  $\al=1$. 
Solid line -  exact values of $e(x)$, dash line - series expansion according to
 Eq. (26). Abscissae are values of running relative equatorial radius $x=r_{eq}/R_{eq}$, 
ordinates are values of ellipticity $e(x)$.} \end{figure}

\begin{figure} \includegraphics[scale=.6]{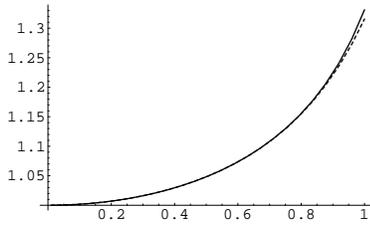}
\caption{Distribution of ellipticity within the star for  $\al=2$. 
See caption to Fig. 4.} \end{figure}

The values of the surface-to-central ellipticity ratio
of ellipticity $e(1)/e(0)$ are $1.6608$ and $2.0488$ at $\al$ equal to $1$ and $2$, 
respectively. In the theory of (slowly) rotating stars the ratio of centrifugal
acceleration $w^2\,$ to the gravitational acceleration $G\,M/R$ at the equator,
$m={w^2 R^3\over G\,M}$ is introduced. Also, the ratio $m$-to-$e(R)$ ratio is introduced,
which in the terms of functions $e(x)$ and $e'(x)$ is ${m\over e}={2\over 5}\,(2+{e'(1)\over
e(1)})$. For the model in question, we have 
$\al=1:\,\,e(1)=1.6608,\,e'(1)=2.0488,\,\, m/e=1.2936,$ and 
$\al=2:\,\,e(1)=1.3312,\,e'(1)=1.3785, \,\,m/e=1.2142$. 

\subsection*{Pade Approximants for the ellipticity distribution}

For the ellipticity distribution inside the star, using the 
series expansion (26), we have Pade(2,2) approximant

\be \label{pad2} \al=2:\quad e_{Pade}(x)={5\,\left( 25025 - 14036\,x + 1138\,{x^2}
\right) \over 7\,\left( 17875 - 13090\,x + 1729\,{x^2} \right)}.\ee

Even at the surface this expression is very accurate:
$e_{Pade}(1)/e_{calc}(1)=0.99891$.  In fact (\ref{pad2}) gives
the analytical solution for the Clairaut's equation for $\al=2$ model.

For the $\al=1$ model Pade(2,2) is also rather accurate:
  
\be \label{pad1} e_{Pade}(x)=
{672 - 448\,x + 41\,{x^2}\over 
   7\,\left( 96 - 88\,x + 15\,{x^2} \right)};\quad \al=1,\ee
with $e_{Pade}(1)/e_{calc}(1)=0.99108$.  

\subsection*{Rotation, contraction and mass loss}

 The abovementioned analytical expressions for parameters of star may be applied
to study of evolution of rotating star, or to evaluation of pulsational
periods of star or to any other problem where the simplicity and explicity of
evaluation justify the approximation certain inaccuracy.\\
We consider in this section the evolution of contracting and rotating star
 with mass loss. Let the rotating star be contracting in  such a way that the distribution
of density follows the law (\ref{denlaw}).
At the some moment the condition for mass loss may be reached at the star's equator.
We assume that the star is spherically-symmetric and rotation is steady-state 
(that is angular rotation is constant over the volume of star, $\omega(r)=constant$)
(that may occur at slow contraction and rapid rotation), then we may write the mass loss
condition as: \be \omega^2={G\,M\over R^3}.\ee
The  rate of angular momentum loss is $dL=\omega\,R^2\,dM$, and taking into account
that $L=k\,\omega\,M\,R^2$, ($k$ is a structure-dependent parameter) we have:
\be {M\over M_0}=\biggl({R\over R_0}\biggr)^\beta;\quad \beta={k\over 2-3\,k}.\ee
As a result, a value of the final mass of the star depends only on parameter $k$,
(see also \cite{AC}). The values of $k$ for polytropes, white dwarfs and
"stepenars" are given in [5 - 8]. For the star with the density distribution
(\ref{denlaw}):

\be k={3+\al\over 5\,(5+\al)};\ee

note that in Eq. (4) the moment is central, and axial moment is $I_{axis}=1/3\,
I_{center}$.\\
At $\al=.06$ that corresponds, by value of ratio $\roroc$, to $n=3$ polytrope, we have
$k\simeq .012$ and $\beta\simeq .078$. Rotation-induced distortion of figure from
sphere leads to $\beta \simeq .02$ \cite{AC}. For the star in the pre-white dwarf stage, a
value of $\beta$ should be even less as the star in pre-white dwarf or pre-neutron star stage is
a (very) hot star with elusive extent envelope and with small dense core (and with very
 small value of $k$). The envelope contains the large rotational momentum and the small
mass therefore the mass loss per unit momentum loss is very small for a such  star.\\
At the other hand for the homogeneous star, $k=.2$, $\beta=1/7$ that is 7 times the
value of $n=3$ polytrope. Therefore if the pre-white dwarf star would be more homogeneous
 then at contraction the rotating star could lose more
 mass. However even in this case to reach a sizable value of mass loss at the
pre-white dwarf stage, the star should have the rotational momentum by order of
magnitude larger that the main sequence star, and for neutron star the difficulty
is even larger. The presence of factors leading to the more homogeneous structure
 could lead to reducing the difficulties in explaining the origin of white dwarfs and neutron 
stars by the mass loss from ordinary stars.

\subsection*{Pulsational periods of pre-white dwarf stars}

Ledoux and Pekeris \cite{LePe} using the energy method obtained the
following formula for the frequency of the fundamental mode of the adiabatic radial
pulsations of the spherically-symmetric star:

\be\label{LP} \sigma^2=(3\,\overline\gamma - 4){-W\over I}.\ee

Using the formulas (4, 5, 12) we obtain

\be \label{LPal} \sigma^2={32\,\pi\over 3}{(1+\al)(4+\al)\over
(5-\al)(5+2\,\al)}\,G\,\overline\rho.\ee

At $\overline\rho\simeq 10^4$ (a star with solar mass and radius $\simeq 1/20$ the solar
radius) and $\al=1$, we have $\sigma\simeq\,8.9\,10^2$, or for the
pulsational period  $P=2\,\pi/\sigma\simeq 70 s$. \\
Such and even larger periods of light variations occur in HL Tau stars, G 44-32 and
R 548 \cite{WaNa}, \cite{LaHe} which are apparently at the late stages of their evolution,
and probably - at pre-white dwarf stage. Recent discussion of pre-white dwarf stars see 
in \cite{Sean}. \subsection*{Radial pulsations}
The differential equation for the adiabatic small radial pulsations of
the spherically-symmetric star is:

\be \label{puls}
{d\over dr} \biggl(r^4\,\gamma\, P \, {d\,y\over d\,r}\biggr)+
y\biggl\{\sigma^2\, \rho\,r^4+r^3\,{d\over
d\,r}[(3\,\gamma-4)\,P]\biggr\}=0;\ee 

here $y=\delta\,r /r$ is the ratio of the radial displacement to the
radius, and $\sigma$ is the frequency of
the pulsations. At the center (at $r=0$), we have condition $\delta\,r=
r\,y=0$, at the surface, $r=R$, we require that amplitude is finite. We
consider here the  case of $\alpha=2$ when Eq. (\ref{puls}) is reduced to
the form:

\be \ba \label{puls2} 2\,y(x)\,x\,(19 - \Sigma^2 - 15\,x^2) + 
  2\,y'(x)\,(-10 +29\,x^2 - 15\,x^4 ) + 
  y''(x)\,(-5\,x +8\,x^3 - 3\,x^5)=0;\ea\ee

here $x=r/R$, and $\Sigma^2$ is dimensionless value, expressed in units
of ${P_c\,\over\rho_c\,R^2}={2\,\pi\over 3}\,G\,\overline \rho$, see Eqs. (2,11).\\
Equation (\ref{puls2}) can be solved in terms of series expansion, and the
recurrence relation can be written down for coefficients of the power series
which may contain only even powers of $x$. Note that Eq. (\ref{puls2}) is
the linear differential equation and the function $y$ is defined up to an arbitrary
factor, so we may take $y(0)=1$. Then the first terms of the series expansion at the
center are:

\be y=1+{19-\Sigma^2\over 25}\,x^2+{490 - 104\,\Sigma^2 + \Sigma^4\over 1000}\,x^4+
{2670 - 12610\,\Sigma^2 + 239\,\Sigma^4 - \Sigma^6\over 60000}\,x^6.\ee

At the surface ($x=1$) we have the series expansion 

$$ y(x)=y(1)\,[1+({\Sigma^2\over 4}-1)\,(1-x)+{112 - 2\,\Sigma^2 +
\Sigma^4\over 48}\,(1-x)^2].$$

For arbitrary $x$ the Eq. (\ref{puls2}) may be solved numerically. We calculated
the four lower modes of radial pulsations and found for the corresponding eigenvalues 
 $\Sigma^2$ the values $10.325,\, 39.083,\,81.04$, and $136.1$, see Fig. 6.

\begin{figure}
\includegraphics[scale=.6]{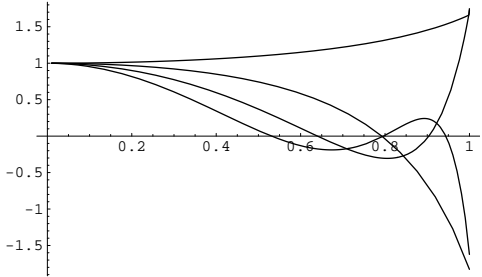}
\caption{Eigenfunctions for 4 lower modes of radial adiabatic pulsations
 as functions of $x=r/R$ for the star model with $\al=2$, see Eq. (35). 
Curves from the
lowest (fundamental) mode (no nodes) to upper mode (3 nodes) correspond  
to values of $\Sigma^2$ equal to $10.325,\,39.083,\,81.04$ and $136.1$,
respectively .} \end{figure}

 The lowest eigenvalue
corresponds to $\sigma^2=10.325*2/3\,\pi\,G\,\overline\rho=
6.88\,\pi\,G\,\overline\rho=$, that only slightly 
differs from value obtained by energy method, $64\,\pi/9\, G\, \overline
\rho=7.1\,\pi\,G\,\overline\rho$, see Eq. (\ref{LPal}) with $\alpha=2$. Note that
the evaluation of $\sigma$ given by Eqs. (\ref{LP}, \ref{LPal}) for the
frequency of the fundamental mode is to be either equal to or greater  
than the true value  \cite{LePe}.\\
The evaluation of pulsational frequency of the lowest mode can also be
obtained directly 
 from Eq. (\ref{puls2}), if we put $y=const$ (which is not too rough
assumption for the fundamental mode eigenfunction, see Fig. 6), then we have 
\be \Sigma^2 =19-15\,x^2.\ee 
We can take averaged value of  right side of this equation with weight $x^2$:
\be \Sigma^2={1\over \int_0^1\,x^2\,dx}\,\int_0^1(19-15\,x^2)\,x^2\,dx=10,\ee
which is very close to the  "exact value" $10.325 $.

\subsection*{Conclusion} 

In conclusion, we present here the analytical model density
distribution which allows to evaluate qualitatively and quantitavely  the many important
characteristics of the  star including the ellipticity distribution within the
rotating star, the pulsational periods and the mass loss at the
contraction stages of evolution.
The certain inaccuracy of the approximation is the modest price paid for the
large simplicity and explicity of the model. Surely, as the first approximation 
or at least as the pedagogical tool the  model is of some interest. 

 \end{document}